# Lateral vortex motion in highly layered electron-doped superconductor $Nd_{2-x}Ce_xCuO_4$.

O.E. Petukhova[1], M.R. Popov[1], A.S. Klepikova[1], N.G. Shelushinina[1], A.A. Ivanov[2], T.B. Charikova[1]

[1]M.N. Mikheev Institute of Metal Physics, Ural Branch, Russian Academy of Sciences, Ekaterinburg, 620108, Russia

[2]National Research Nuclear University MEPhI, Moscow, 115409, Russia

**Abstract**

The carrier transport and the motion of a vortex system in the electron-doped high-temperature superconductors $Nd_{2-x}Ce_xCuO_4$ in underdoped and optimally doped ($x$ = 0135, 0.145, 0.15) regions, in the area of the evolution from antiferromagnetic to superconducting order were investigated. To study the anisotropy of the transport properties of highly layered NdCeCuO system we have synthesized $Nd_{2-x}Ce_xCuO_4/SrTiO_3$ epitaxial films of three types with different orientations of the *c*-axis and conductive $CuO_2$ layers relative to the substrate. Such a set of samples allowed us to study the processes of both standard (in the $CuO_2$ layers) and lateral (across the $CuO_2$ layers) carrier transfer in the normal and the mixed states of a superconductor. In a flux-flow regime, in magnetic field ***B***, the dynamics of Abrikosov (***B***∥*c*-axis) and Josephson (***B***∥*ab*-plane) vortices are thoroughly investigated and analyzed which is perspective for scientific purposes and for practical applications in measurement technology.

**Introduction**

Resonators, nano-SQUIDs, voltage standards and energy storage devices based on layered think and thick films are increasingly used in measurement technology [1, 2]. The study of transport properties of layered epitaxial films of electron doped superconductors in crossed electric and magnetic fields as new objects for use in electrical engineering is a necessary part of scientific research.

From the beginning of the discovery of high-temperature superconductors (HTSC), the Hall effect in these compounds has been actively studied. In the mixed state of HTSC, as in conventional type II superconductors one, their transport properties in the external magnetic field are determined by the motion of Abrikosov vortices [3]. However, unlike isotropic superconductors, HTSC have a strong anisotropy of their properties. This anisotropy is caused by a layered crystal structure formed by the conductive $CuO_2$ layers (*ab*-planes) with spacing between them in *c*-direction. Depending on the applying direction of the external magnetic field there are different situations:

a. *The magnetic field is parallel to the c-axis.* In this case, vortices in the superconducting layer distribute in a regular triangular lattice as in the conventional superconductors. However, the coherence length along the *c*-axis is smaller than the plane spacing, and the description of the flux vortices must be modified to take into account the fact that they are localized within the planes. Thus, the result is a stack of "pancakes"[4]. The flux line is connected weakly to each other with pancake vortex in each layer. Under the action of the transport current, the pancake vortices will move along the *ab*-planes producing dissipation. Since the density of pancake vortices is proportional to $B/B_{c2}$, $B_{c2}$ being the second critical field, the corresponding flux flow resistivity is also proportional to $B/B_{c2}$ [5]. Dissipation by flux flow does not begin immediately as the vortices enter type II superconductors due to a pinning force opposite to the Lorentz force. Such vortex pinning sites are provided by defects in superconductors, which act as energetically favorable sites at which a flux line can fixed [6]. However, even if the average pinning force remains stronger than the Lorentz force and flux flow is prevented, there is a dissipation caused by thermal fluctuations. One or more flux lines may jump from one pinned configuration to another, overcoming the energy barrier by thermal activation, thus the flux creep occurs.

b. *The magnetic field is parallel to conductive ab -planes*. In this case, vortices appear to be as in Josephson junctions of superconductor/dielectric/superconductor structures because the layered structure of HTSC forms internal Josephson junctions [7, 8]. The Josephson vortices, each of which carries a magnetic flux quantum and whose center is between the superconducting layers have no normal core and thus do not strongly suppress the order parameter in the adjacent superconducting planes [9]. The motion of Josephson vortices in HTSC is quite different from that of pancake vortices. Josephson vortices can easily move along the superconducting layers, but not perpendicular to them. This is so-called intrinsic pinning of Josephson vortices [8]. For the long Josephson junctions in a parallel magnetic field there is a specific magnetic field, the crossover field, at which the vortices begin to overlap, forming a triangular lattice of Josephson vortices [9]. The movement of the Josephson lattice under the action of transport current leads to the appearance of the flux-flow resistivity. In this situation, the intrinsic interlayer pinning is significant [10]. Often such objects are considered theoretically as layered superconducting structures with weak Josephson interaction [11]. Moreover, the motion of vortices through Josephson barriers was discussed long before the appearance of HTSC [12].

c. *For fields at intermediate angles*, the vortex can be described as a combination of pancake vortices in the *c*-direction (confined within $CuO_2$ layers) connected by Josephson vortices in the *ab*-plane [13].

Experimental investigations of the P.H.Kes group on the conventional superconductors such as amorphous $Nb_3Ge$ [14] and $NbN/Nb_3Ge$ double layer [15] and on hole-doped HTSC $Bi_2Sr_2CaCu_2O_8$ and $YBa_2Cu_3O_7$ [16] have shown that there are some particularities of the vortex dissipation: oscillations and fluctuations [15, 17]. In [18] the authors have associated negative anomaly of the Hall effect in $YBa_2Cu_3O_7$ with the striking sign difference between Hall response of vortices that lie parallel to the layers and perpendicular to it.

For electron-doped HTSCs, the situation is complicated by the fact that antiferromagnetic (AFM) phase persists to much higher doping levels than in the hole-doped systems, and thus the underdoped region of the superconducting (SC) phase is hidden by AFM ordering and/or by the coexistence of spin density waves with the onset of SC ordering [19]. In recent years an interest on the study of transport properties in *n*-type cuprates sharply increased [19-25] including studies on $Nd_{2-x}Ce_xCuO_4$ films with different orientations [21, 22] and $Nd_{2-x}Ce_xCuO_{4\pm\delta}$ ultrathin crystalline films [23-25].

In this work, we analyzed carrier transport and the motion of a vortex system in the electron-doped HTSC $Nd_{2-x}Ce_xCuO_{4+\delta}$ in underdoped and optimally doped ($x$ = 0135, 0.145, 0.15) regions, in the area of the evolution from AFM to SC order. We have made the comparison of the processes developing in the conducting $CuO_2$ layers and across the layers.

**Materials and methods**

For this purpose, we have synthesized by pulsed laser deposition $Nd_{2-x}Ce_xCuO_{4+\delta}/SrTiO_3$ epitaxial films with optimal oxygen content $\delta$=0 and $x$ = 0.135, 0.145 and 0.15 of three types [26]:

I - films with standard orientation (001) where the *c*-axis of the $Nd_{2-x}Ce_xCuO_{4+\delta}$ lattice is perpendicular to the $SrTiO_3$ substrate plane; to measure the longitudinal $\rho_{xx}^{ab}$ and Hall $\rho_{xy}^{ab}$ resistivities in $CuO_2$ layers.

II- films (1$\bar{1}$0) where the *c*-axis of the $Nd_{2-x}Ce_xCuO_{4+\delta}$ lattice is directed along the long side of the $SrTiO_3$ substrate; to measure the longitudinal $\rho_{xx}^{c}$ resistivity between (across) $CuO_2$ layers.

III - films (1$\bar{1}$0) where the *c*-axis of the $Nd_{2-x}Ce_xCuO_{4+\delta}$ lattice is directed along the short side of the $SrTiO_3$ substrate; to measure the Hall resistivity $\rho_{xy}^{c}$ between (across) the $CuO_2$ layers.

In the process of pulsed laser deposition, an excimer KrF laser was used with a wavelength of 248 nm, with an energy of 80 mJ/pulse. The energy density at the target surface is 1.5 J/cm². The pulse duration was 15 ns, the repetition rate of pulses was from 5 to 20 Hz. Further, the synthesized films were subjected to heat treatment (annealing) under various conditions to obtain samples with a maximum superconducting transition temperature. X-ray diffraction analysis (Co-K radiation) showed that all films were of high quality and were single crystal.

The optimum annealing conditions were as follows:
- for the composition $x$ = 0.15 ($T^c_{onset}$ = 23.5 K, $T_c$ = 22 K) - $t$ = 60 min., $T$ = 780$^0$C, $p$ = 10$^{-5}$Torr;
- for the composition $x$ = 0.145 ($T^c_{onset}$ = 15.7 K, $T_c$ = 10.7 K) - $t$ = 60 min., $T$ = 600$^0$C, $p$ = 10$^{-5}$Torr;
- for the composition $x$ = 0.135 ($T^c_{onset}$ = 13.7 K, $T_c$ = 9.6 K) - $t$ = 60 min., $T$ = 600$^0$C, $p$ = 10$^{-5}$Torr.

The thickness of the films was $d$ = 140-520 nm.

For correct measurement using the standard four-probe method, the geometry of the samples was six-contact Hall bar. In the process of the Hall effect measure in superconductors, there are several measurement errors, which in magnitude can be larger than the measured signal:

a. Misalignment voltage connected with imperfect sample geometry. These voltages are not magnetic field dependent and can be compensated by use field reversal to remove by subtraction.

b. Thermal electric voltage due to temperature effects. These voltages are not current dependent and can be compensated by use current reversal to remove by subtraction.

Thus, four measurements (for two directions of current and voltage) were performed to separate the Hall voltage from the misalignment voltage and thermal electric voltage.

The temperature and magnetic field dependences of the longitudinal, $\rho_{xx}(B,T)$, and Hall, $\rho_{xy}(B,T)$, resistivity for all the types of $Nd_{2-x}Ce_xCuO_{4+\delta}/SrTiO_3$ films were investigated in the Quantum Design PPMS 9 and in the solenoid "Oxford Instruments" (Center for Nanotechnologies and Advanced Materials, IFM UrB RAS). The electric field was applied parallel to the $SrTiO_3$ substrate plane. The external magnetic field $\boldsymbol{B}$ was directed perpendicular to the plane of the $SrTiO_3$ substrate. Depending on the type of films, we were able to measure the magnetic field dependences of the resistivity in the conducting $CuO_2$ layers and between (across) layers (along the $c$-axis) in crossed electric and magnetic fields.

**Experimental results and discussion**

Flux flow in layered superconductors is a rich and complicated phenomenon. The fundamental difference with respect to homogeneous isotropic and anisotropic superconductors is a strong interaction between vortices and the crystal structure itself known as "intrinsic pinning". This introduces new features to the vortex dynamics.

An important effect specific for layered systems, namely, an interaction of the order parameter with the underlying crystalline structure can be considered on the example of vortices, which are aligned parallel to the superconducting planes. The interaction with the inhomogeneous layered environment gives rise to the intrinsic pinning: the vortex energy becomes dependent on the vortex position with respect to the planes and produces a pinning force which tries to keep vortices in between the superconducting layers.

A schematic image of the investigated samples, showing the location of the layers and the shape of device for each type of the film is presented on Fig. 1.

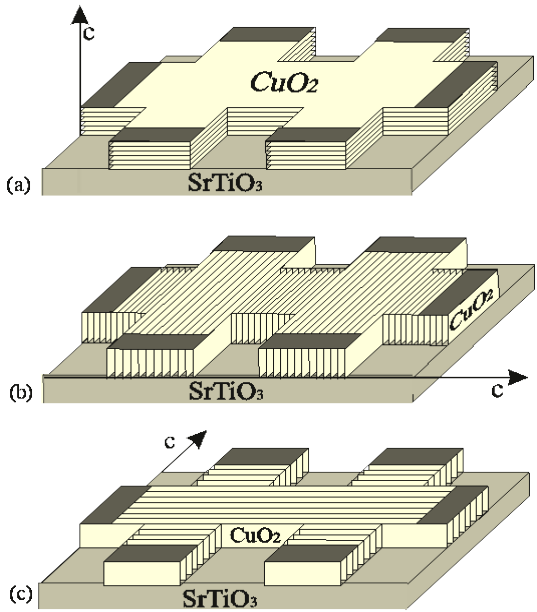

Fig. 1.The orientations of conductive $CuO_2$ layers relative to the substrate $SrTiO_3$ for the films of different type: I – type films (a); II – type films (b) and III – type films (c).

In-plane $\rho_{xx}^{ab}(B)$ ( I – type films, Fig. 1a) and out-of-plane magnetoresistivity $\rho_{xx}^{c}(B)$ ( II-type films, Fig. 1b) with different doping level $x$=0.135, 0.145, 0.15 and at different temperatures were investigated. Figures 2a, b show the magnetic field dependencies of the in-plane resistivity $\rho_{xx}^{ab}$ and out-of-plane resistivity $\rho_{xx}^{c}$ at the temperature T = 4.2K.

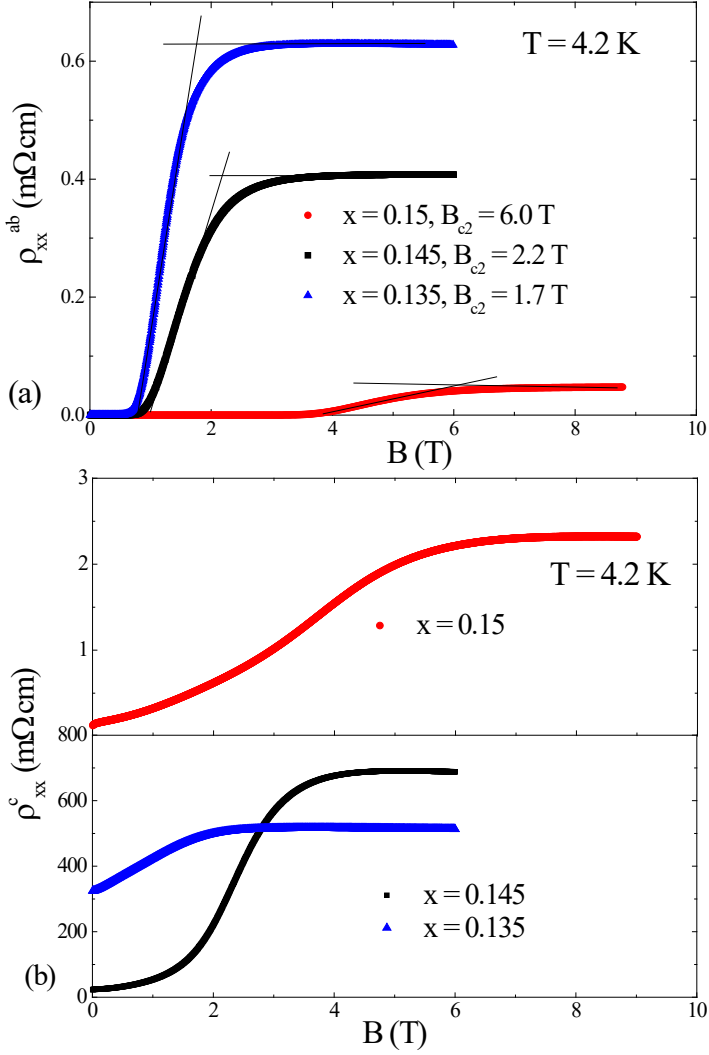

Figure 2. The magnetic field dependencies of the in-plane resistivity $\rho_{xx}^{ab}(B)$ (a) and out-of-plane resistivity $\rho_{xx}^{c}$ (b) for $Nd_{2-x}Ce_xCuO_4/SrTiO_3$ films with $x$=0.135, 0.145, 0.15 and optimal annealing at $T$=4.2 K. The location of the layers and the shape of the sample matches on Figs 1a,b.

Strong anisotropy of the resistivity in the normal state (at $B>B_{c2}$) is observed along and across the conductive layers: the ratio $\rho_{xx}^{c}(B)/\rho_{xx}^{ab}(B) \sim 10^2$ at $B$ = 8T for $x$=0.15 and $\rho_{xx}^{c}(B)/\rho_{xx}^{ab}(B) \sim 10^3$ at $B$ = 6T for $x$=0.145 and 0.135 (Fig. 2).

As can be seen, with an insignificant increase of the doping level from $x$ = 0.135 to $x$ = 0.15 leads to decrease by an order of magnitude of the in-plane normal state resistivity in the magnetic fields higher that upper critical field and by more than two orders of the out-of-plane normal state resistivity. Moreover, the out-of-plane resistivity does not disappear even at weak magnetic fields.

In the mixed state, the behavior of the transverse resistivity from the magnetic field differs from the behavior of the resistivity along the conductive layers. In $\rho_{xx}^{ab}(B)$ dependence (Fig. 2a) there is a pronounced boundary of a transition from the SC ($\rho = 0$) to the resistive state ($\rho \neq 0$) at the vortex-depinning field $B_{dp}$. It is a field in which the pancake vortices, having normal cores on the layers, start to move along them, perpendicular to a transport current, thus producing dissipation. At $B_{dp} < B < B_{c2}$ good agreement was found with the ordinary behavior $\rho_{xx}{}^{ab}(B) \sim B/B_{c2}$ [27].

On the other hand, for out-of-plane magnetoresistivity, $\rho_{xx}{}^c$, an overcoming of the "intrinsic pinning" of Josephson vortices, aligned parallel to the superconducting planes, occurs gradually with the dependence $\rho_{xx}{}^c(B) \sim B^2$ when going into the resistive state (Fig. 2b).

The theoretical conceptions for dynamics of Josephson vortices in a layered superconductor are developed by Koshelev [9]. A field dependence of the flux-flow resistivity $\rho_{ff}(B)$, for the case of dominating in-plane dissipation channel is obtained in the form:

$$\rho_{ff} = \frac{B^2}{B^2+B_\sigma^2}\rho_c \quad (1)$$

with $\rho_c$ being the flux-flow saturation resistivity along the $c$-axis, $B_\sigma = \sqrt{\frac{\sigma_{ab}}{\sigma_c}}\frac{\Phi_0}{\sqrt{2\pi}\gamma^2 s^2}$ where $\sigma_{ab}$ is the in-plane quasiparticle conductivity, $\sigma_c$ is the c-axis component of conductivity, $\gamma$ is the anisotropy of London penetration depth, $s$ is the interlayer spacing and $\Phi_0$ is the magnetic flux quantum.

Thus, for strong in-plane dissipation the indicated dependence $\rho_{ff}(B)$ should have pronounced upward curvature at $B < B_\sigma$ and approaches $\rho_c$ as the field $B \to B_\sigma$, as shown schematically in Fig.1 of Ref. [9]. In general, the behavior of $\rho_{xx}{}^c(B)$ in the mixed state of our films is in accordance with Eq. (1) (see Fig. 2b).

Investigation of the Hall effect in the normal and mixed states gives important information about the vortex dynamics for high-$T_c$ superconductors. The magnetic field ***B*** penetrates into a type-II superconductor by quantum vortices [3-5]. The vortex motion along the Lorentz force (perpendicular to the transport current ***j***) generates a dissipative field (***E***||***j***) and leads to the flux-flow resistivity. On the other hand, the vortex motion along the direction of transport current will result in the Hall electric field (***E***$_H$⊥***j***, ***B***). Thus, the mixed state Hall effect measurement is the useful method to study vortex dynamics in the investigated systems.

We performed a comparative study of the magnetic field dependencies for the in-plane Hall resistivity $\rho_{xy}{}^{ab}(B)$ (measured on I – type films) and the Hall resistivity between CuO$_2$ layers $\rho_{xy}{}^c(B)$ (measured on III – type films). The results for films with $x$ = 0.135, 0.145, 0.15 at $T$ = 4.2 K are presented on Fig. 3.

When measuring in a standard situation on films of the I-type, we have ***j***||*ab*, ***B***||*c* (see Fig. 1a) and the Hall field, ***E***$_H$⊥ (***j***, ***B***), lies in the *ab*-plane, i.e. in mixed state we are dealing with Abrikosov vortices moving in *ab*-planes. As in a case of $\rho_{xx}^{ab}$, we see a pronounced vortex-depinning field $B_{dp}$ ($\rho_{xy}^{ab} = 0$ for $B<B_{dp}$), then a narrow region of a mixed state (this region is described in detail in our previous works, see, for example, [22, 28]). For the film with optimal doping, we can see a sign change of the Hall resistivity in the mixed state that reflects the Fermi surface transformation [22].

The in-plane Hall resistivity in normal state has the standard magnetic field dependence $\rho_{xy}{}^{ab}(B) = R_H B$, the sign of the Hall effect is negative for all the three doping levels. By the value of the Hall coefficient, we can estimate the concentration of carriers (electrons) in the normal phase: $n = 3.05 \cdot 10^{21}$ 1/cm$^3$ ($x = 0.135$); $n = 6.25 \cdot 10^{21}$ 1/cm$^3$ ($x = 0.145$); $n = 2.16 \cdot 10^{22}$ 1/cm$^3$ ($x = 0.15$).

Another situation is in the films of the III – type (see Fig. 1c) : with $j \| ab$, $B \| ab$ and $j \perp B$ the Hall field $E_H \perp (j, B)$ drives the motion of carriers across the CuO$_2$ planes, along the $c$-axis. In the mixed state, it corresponds to a flux-flow of Josephson vortices with the intrinsic pinning. For $x = 0.135$ and $x = 0.15$ in weak magnetic fields there is no region where $\rho = 0$, as in the case of $\rho_{xx}^c$: there is a maximum on the $|\rho_{xy}^c(B)|$ in the mixed state, the sign of the Hall effect is positive for $x = 0.135$ and negative for $x = 0.15$ (see Fig. 3b). For $x = 0.145$ $\rho_{xy}^c = 0$ up to $B \approx 3T$ and then some peculiarities of a maximum – minimum type is observed with a change in the sign of $\rho_{xy}^c(B)$ (see Figs. 4a and 6a below).

In the normal state, the Hall resistivity between CuO$_2$ layers is practically independent on magnetic field.

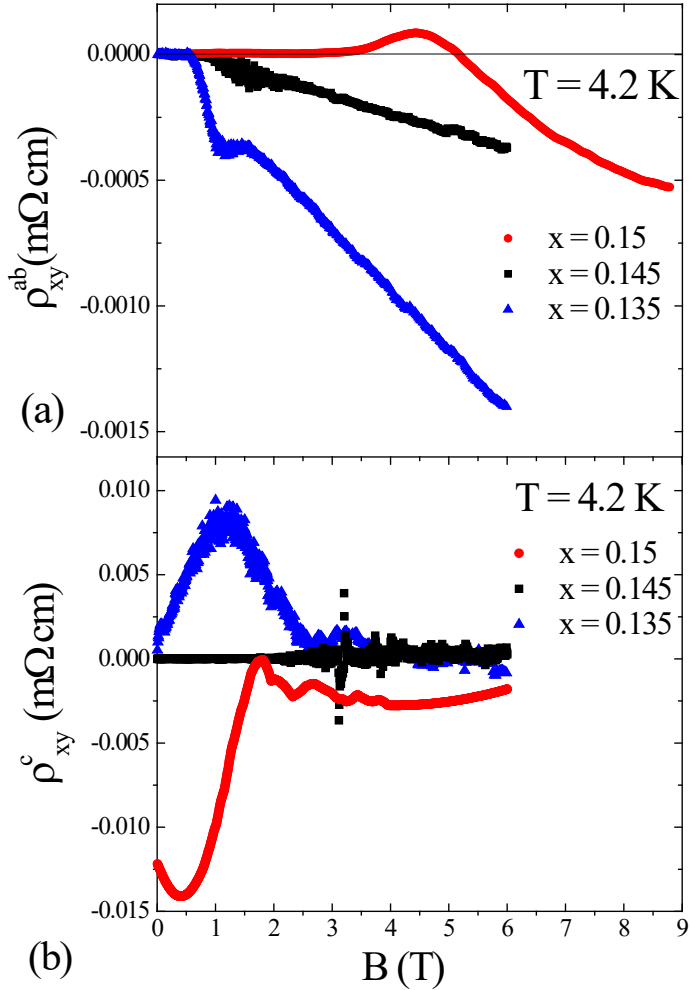

Figure 3. The magnetic field dependencies of the Hall resistivity for Nd$_{2-x}$Ce$_x$CuO$_4$/SrTiO$_3$ films with $x=0.135, 0.145, 0.15$ at $T= 4.2$ K: (a) – in-plane Hall resistivity, (b) – Hall resistivity between CuO$_2$ layers.

It was argued [29-31] that in the mixed state of a type II superconductor, because of the difference of the chemical potential in a superconducting versus normal state, the vortex cores may become charged. The difference $\delta n$ between the electron density at the center of the vortex core and that far outside the vortex causes the additional (topological) flux-flow contribution to the Hall conductivity $\sigma_{xy}^{ff} = -e\delta n/B$.

Thus, in a vicinity of the transition from the superconducting to the resistive state the Hall conductivity should strive to a large positive or negative value with decreasing field. Such $\sigma_{xy}(B)$ dependencies repeatedly observed in the mixed state both of *p*-type oxide superconductors and of Fe-based ones (see [28, 32] and references therein).

In [1, 33, 34] vortex dynamics is considered within the framework of the time dependent Ginzburg-Landau (TDGL) model. A modification of the TDGL model is considered which allows to account for the flux-flow Hall effect. According to [1, 33, 34], an anomalous contribution into Hall effect in a flux-flow regime due to the motion of vortices parallel to the transport current can be represented as

$$\sigma_{xy} = -e\delta n/B, \quad \delta n = \left(\frac{1}{\lambda}\frac{dv}{d\mu}\right)\Delta^2, \tag{2a}$$

where $\delta n$ is the "virtual" variation in the electron density caused by the change in the electronic spectrum after transition into the superconducting state, $\lambda, v, \mu,$ and $\Delta$ being the penetration length, density of states in the normal phase, chemical potential and the order parameter magnitude (superconducting gap), respectively. In a superconductor, there is no real electron density change because of the charge neutrality: all the variations are compensated by the corresponding variations in the chemical potential.

The magnitude of $\delta n$ can be expressed also via the experimentally accessible quantities [30]:

$$\delta n = -\frac{H_c^2}{4\pi}\frac{\partial \ln T_c}{\partial \mu}, \tag{2b}$$

where $H_c$ is the critical magnetic field and $T_c$ is the critical temperature. Thus, the sign of the Hall effect in the mixed state depends on the details of the band structure (see also Ref. [35]).

The dependences of $\rho_{xy}^c(B) \sim \sigma_{xy}(B)$ in the mixed state, observed by us (see Fig.3b), can be associated with the anomalous vortex contribution into the flux-flow Hall effect with $\delta n = \delta n_h > 0$ for $x = 0.135$ and $\delta n = \delta n_e < 0$ for $x = 0.15$ in Eq.(2a), $\delta n_e$ and $\delta n_h$ being the electron- and hole-like parts of $\delta n$. For $x = 0.145$ it can be assumed, that $\delta n = \delta n_h + \delta n_e$, here with the electron and the hole contributions cancel each other, $\delta n \cong 0$, at $B < 3T$ and then their existence leads to a nonmonotonic $\rho_{xy}^c(B)$ dependence with a change in the sign of the Hall effect (see Figures 4a and 6a). The sign of $\sigma_{xy}$ in the mixed state may be different for a complicated Fermi surface which has electron-like and hole-like parts [35].

We now carry out a comparative analysis of the magnetic field dependences of the longitudinal and Hall resistivity, both of which correspond to the motion of carriers across the $CuO_2$ layers: $\rho_{xx}{}^c(B)$, measured on II-type films, and $\rho_{xy}{}^c(B)$, measured on III-type films, for samples with $x=0.135$ and $x=0.145$. Figure 4 clearly shows that the nonmonotonic $\rho_{xy}{}^c(B)$ dependencies are inherent just to the mixed state regions, i.e., as suggested above, reflect the dynamics of the Josephson vortices.

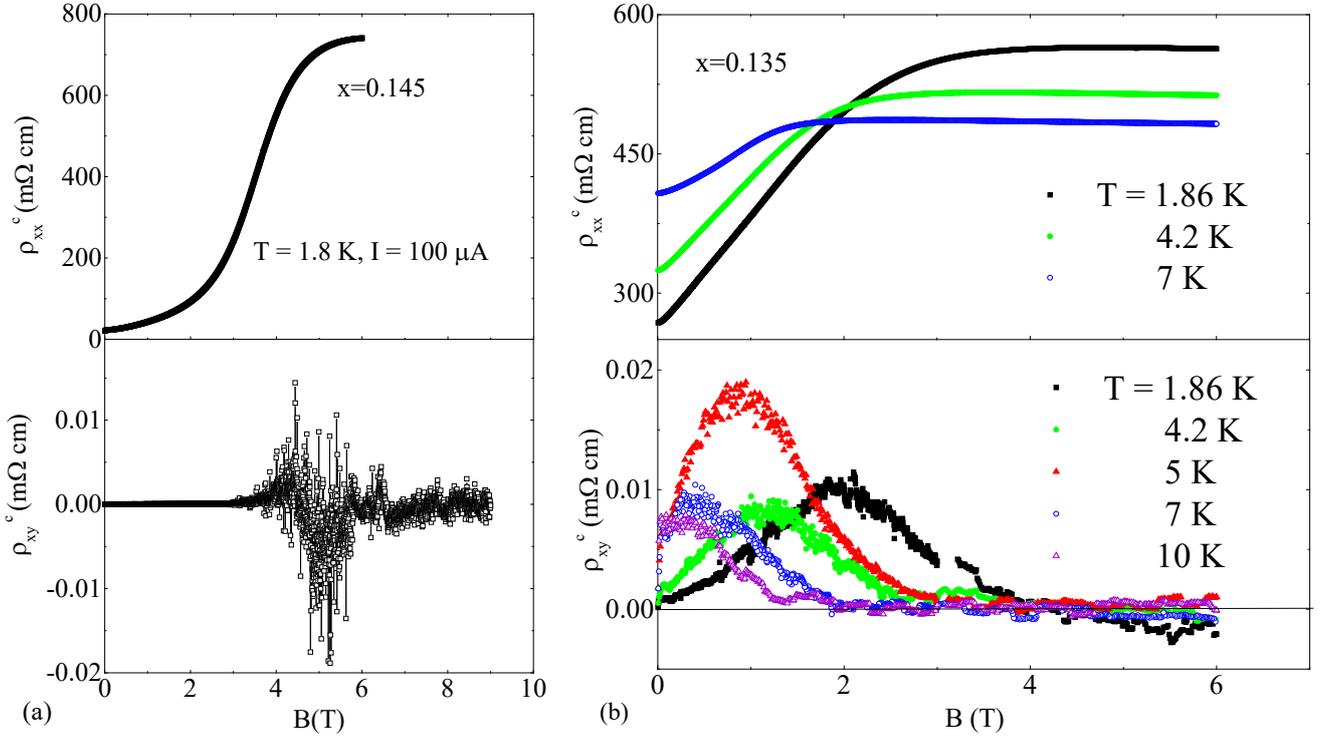

Figure 4. Magnetic field dependencies of the resistivity and Hall resistivity between CuO$_2$-planes for Nd$_{2-x}$Ce$_x$CuO$_{4+\delta}$/SrTiO$_3$ films with $x = 0.145$ at $T=1.8$ K at different temperatures (a) and $x=0.135$ at different temperatures (b).

In the situation when the magnetic field and vortices are aligned parallel to the layers and the transport current flows in the plane of layers the Lorentz force will be directed perpendicular to the layers (the $z$-axis is to be along the crystallographic $c$-direction, the $x$ and $y$-axes lies in CuO$_2$ layer). The vortex line motion equation according to [36, 37] has the form:

$$\frac{\Phi_0}{c} J_{tr}(\omega) \times z_0 = \frac{V_{Lx(\omega)}}{\mu_x} x_0 + \frac{V_{Ly(\omega)}}{\mu_y} y_0, \qquad (3)$$

Where $x_0$, $y_0$, $z_0$ are the unit vectors of the coordinate system, $V_{Lx(\omega)}$ and $V_{Ly(\omega)}$ are components of the vortex velocity, $\mu_x$ and $\mu_y$ are the vortex mobility in $x$ and $y$ directions, the left side of the equation is the Lorentz force ($J_{tr}$ being the transport current). Straight Josephson vortices are aligned parallel to the layers in the magnetic field parallel to CuO$_2$ planes and begun to move in the direction perpendicular to the layers overcoming their own intrinsic pinning (the lateral flux-flow).

The peculiarities of the $\rho_{xy}^c(B)$ behavior, observed for Nd$_{2-x}$Ce$_x$CuO$_{4+\delta}$/SrTiO$_3$ films with $x=0.135$ and $x = 0.145$ in the flux-flow regime are presented in detail in Fig.4: the $\rho_{xy}^c(B)$ behavior has a positive maximum in the first (data at different temperatures are shown) and $\rho_{xy}^c(B)$ has a double sign change in the second one (data on a larger scale are shown). The specific form of the $\rho_{xy}^c(B)$ curves may be associated with the dependences of the $\delta n$ in Eq.(2a) on $B$ and $T$.

A normalized charge density at the vortex center, $\delta n/n$, as a function of magnetic field at different values of $T/T_c$ was calculated in [40]. The field dependence of the charge density at the core center is well described by $B(B_{c2} - B)$ with a peak near $B_{c2}/2$ originating from competition between the increasing magnetic field and the decreasing pair potential. Herewith, maximum $\delta n/n$ value significantly depends on the temperature (see Fig. 3 in [40] for $T/T_c = 0.2$ and $T/T_c = 0.5$), which corresponds to the effect observed by us (see Fig.4b).

Analyzing the field dependence of the Hall resistivity between $CuO_2$ layers at different temperatures for $x = 0.145; 0.135$, we saw exponential increase of the Hall resistivity between $CuO_2$ layers at T = 1.8 K and in the range of the magnetic field $B = (2.5 - 4.4)$ T for the first compound and power dependence of the Hall resistivity between $CuO_2$ layers at low temperatures and at low magnetic fields $\rho_{xy}^c \sim B^\beta$ for the second one (Figure 5). At the temperature $T = 1.86$ K in magnetic fields $B = (0.14 - 0.62)$ T index $\beta$ is equal $\beta = 0.8$, in the range of the magnetic field $B = (0.62 - 1.84)$ T index $\beta$ changes to $\beta = 1.4$; at the temperature $T = 4.2$ K in the range $B = (0.13 - 1.1)$ T index $\beta$ is always $\beta = 0.8$, at the temperature $T = 5$ K in the range $B = (0.1 - 0.7)$ T is getting smaller up to $\beta = 0.6$ and at $T = 7$ K and upper we cannot see any power dependencies of the Hall resistivity between $CuO_2$ layers.

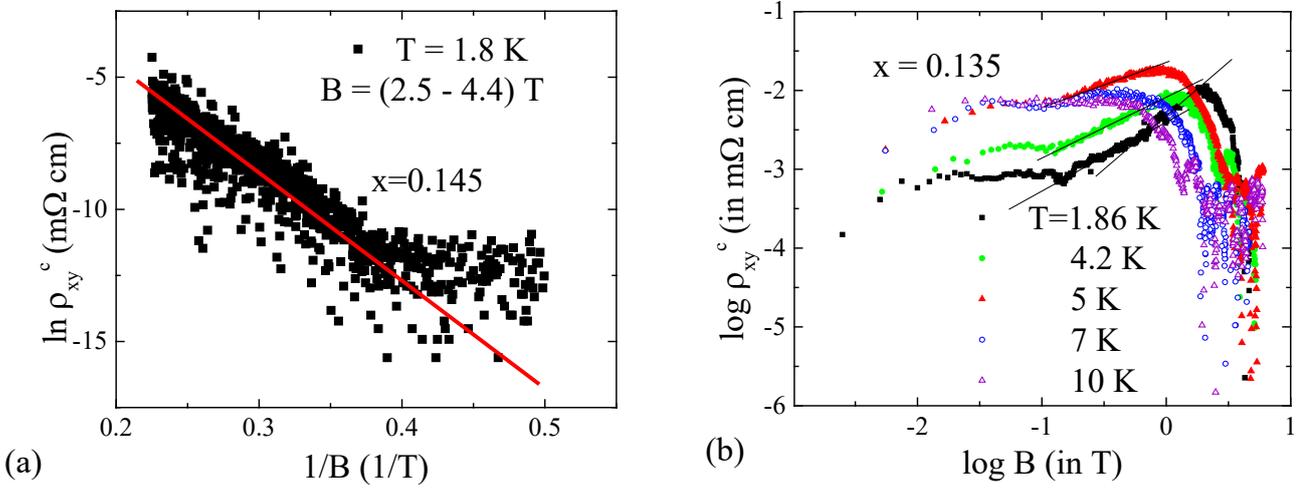

Figure 5. Exponential increase of the Hall resistivity between $CuO_2$ layers in the range of the magnetic field $B = (2.5 - 4.4)$ T for $Nd_{2-x}Ce_xCuO_{4+\delta}/SrTiO_3$ films with $x=0.145$ at $T = 1.8$ K (a); Power dependence $\rho_{xy}^c(B)$ for $Nd_{2-x}Ce_xCuO_{4+\delta}/SrTiO_3$ films with $x=0.135$ at different temperatures - log-log plot (b).

The increase of the magnetic field causes the growth of the vortices number with the orientation of the magnetic moment along the magnetic field and, as a result, the decrease of the vortex-antivortex interaction [38, 39]. We see a change in the index on the power law of the Hall resistivity curves between the $CuO_2$ layers from $\beta = 0.8$ to $\beta = 1.4$ at magnetic field $B = 0.62$T and temperature $T = 1.86$ K. The fitting parameters of the magnetic field Hall resistivity power dependencies were determined (Figure 6b). The change of the power law index of the Hall resistivity curves between the $CuO_2$ layers may indicate the Berezinskii-Kosterlitz-Thouless (BKT) transition that manifested in two-dimensional systems [38, 39]. An important component of the study of the magnetic and temperature dependences of the resistivity and Hall resistance is the question of the possibility of identifying the BKT transition in two-dimensional NdCeCuO films depending on the orientation of the external magnetic field with respect to the conductive layers of $CuO_2$. Some efforts in this direction were made on NbN films in the study of current – voltage characteristics [41].

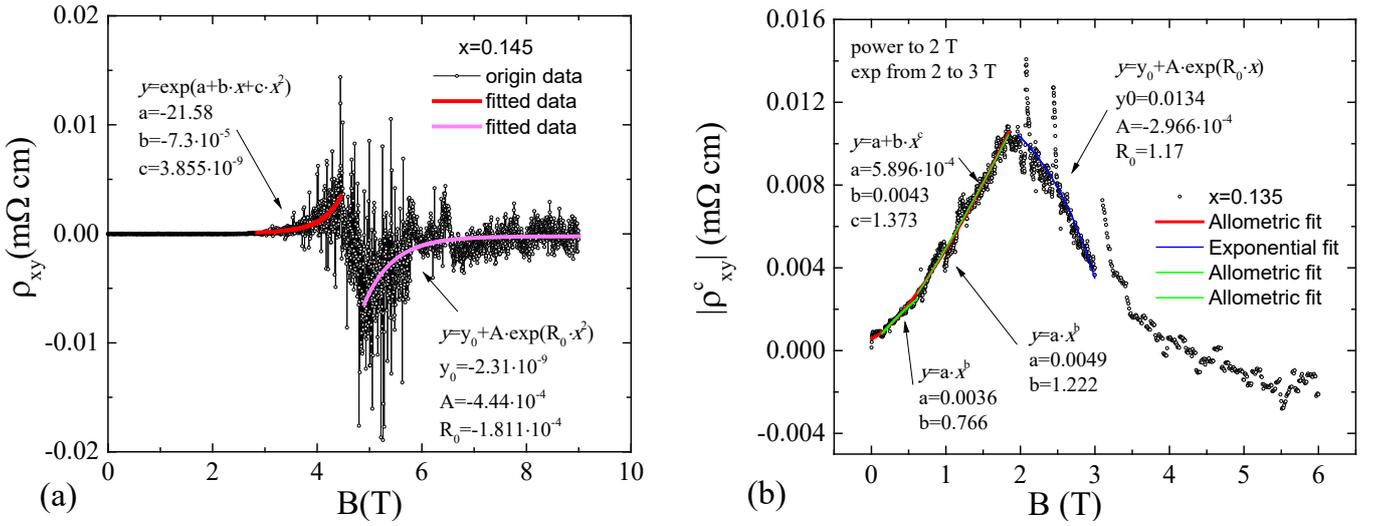

Figure 6. Magnetic field dependence of the Hall resistivity between $CuO_2$ layers for $Nd_{2-x}Ce_xCuO_{4+\delta}$/$SrTiO_3$ films with $x=0.145$ (a); 0.135 (b) at $T = 1.86$ K: experimental data and fitting.

More complicated magnetic field dependence of the Hall resistivity between $CuO_2$ layers for $Nd_{2-x}Ce_xCuO_4$/$SrTiO_3$ films with $x=0.145$ were found (Figure 5a and 6a). We can see double sign change of the Hall resistivity value $\rho_{xy}^c$ in the mixed state at $B = (2.5 – 6.5)$ T. The increase of the magnetic field from $B = 2.5$ T leads to the exponential dependence of $\rho_{xy}^c$ due to increase of the number of the vortexes that appear parallel to $CuO_2$ layers and begin to move between these planes. The change of the Hall resistivity sign of $\rho_{xy}^c$ can testify the existence of the vortexes with the core of the second type of carriers. The presence of the Fermi surface areas of two types of carriers corresponds to this $Nd_{2-x}Ce_xCuO_4$ compound with $x = 0.145$.

**Conclusion**

Thus, the magnetic field dependences of both in-plane and out-of-plane longitudinal, $\rho_{xx}$, and Hall, $\rho_{xy}$, resistivities in underdoped ($x= 0.135$ and $0.145$) and optimally doped ($x= 0.15$) $n$-type layered superconductor $Nd_{2-x}Ce_xCuO_4$ were investigated. The measurements were carried out on $Nd_{2-x}Ce_xCuO_4$/$SrTiO_3$ epitaxial films with different orientations of the $c$-axis and the $ab$–planes relative to the substrate in order to make a comparative analysis of the processes developing in the conducting $CuO_2$ layers and across them. The peculiar properties of $\rho_{xx}(B)$ and $\rho_{xy}(B)$ behavior in the mixed (resistive) state are associated with the lateral motion of the vortex system in crossed electric and magnetic fields.

**Acknowledgments**

The work was carried out within the framework of the state assignment on the subject "Electron" № AAAA-

A18-118020190098-5 with the support of the Russian Foundation for Basic Research (grant № 18-02-00192 and grant № 20-42-660004 with the partial financing under the contract № 50-CO). The authors would like to thank S.M. Podgornikh for support of measurements on PPMS 9.

**Reference**

1 F. W.Carter, T. Khaire, C. Chang and V. Novosad, Low-loss single-photon NbN microwave resonators on Si, Appl. Phys. Lett., 2019, V. 115, P. 092602.

2 Wu Long, C. Lei, W. Hao, W.Qisi, Wo Hongliang, Z. Jun, L.Xiaoyu, Wu Xiaolei and W. Zhen, Measurement of Meissner effect in micro-sized Nb and FeSe crystals using an NbN nano-SQUID, Supercond. Sci. Technol., 2017, V. 30, P. 074011.

3 N. B. Kopnin, Theory of Nonequilibrium Superconductivity, International Series of Monographs on Physics, Oxford University Press, 2009, P. 344.

4 M. Tinkham, Vvedeniye v sverkhprovodimost', M., Atomizdat, 1975.

5 G. Blatter, M.V. Feigel'man, V.B. Geshkenbein, A.I. Larkin and V.M. Vinokur, Vortices in high-temperature superconductors, Rev. Mod. Phys., 1994, V. 66, P. 1125-1388.

6 W. K. Kwok, U. Welp, V. M. Vinokur, S. Fleshler, J. Downey, and G. W. Crabtree, Direct observation of intrinsic pinning by layered structure in single-crystal $YBa_2Cu_3O_{7-\delta}$, Phys. Rev. B, 1991, V. 67, P. 390-393.

7 R. Kleiner, F. Steinmeyer, G. Kunkel and P. Muller, Intrinsic Josephson Effects in $Bi_2Sr_2CaCu_2O_8$ Single Crystals, Phys. Rev. Letters, 1992, V.68, P.2394-2397.

8 M. Rapp, A. Murk, R. Semerad and W. Prusseit, c-Axis Conductivity and Intrinsic Josephson Effects in $YBa_2Cu_3O_7$, Phys. Rev. Letters, 1996, V.77, P.928-931.

9 A. Koshelev, Role of In-Plane Dissipation in Dynamics of a Josephson Vortex Lattice in High-Temperature Superconductors, Phys. Rev. B, 2000, V. 62, P. R3616-3619.

10 B.I. Ivlev and N. B. Kopnin, Flux Creep and Flux Pinning in Layered High-Temperature Superconductors, Phys. Rev. Letters, 1990, V. 64, P. 1828-1830.

11 G. Blatter, B. I. Ivlev, J. Rhyner Kosterlitz-Thouless, Transition in the Smectic Vortex State of a Layered Superconductor, Phys. Rev. Letters, 1991, V. 66, P. 2392-2395.

12 P.Lebwohl, M.J.Stephen, Properties of vortex lines in superconducting barriers, Physical Review, 1967, V.163, P. 376-379.

13 L. N. Bulaevskii, M. Maley, H. Safar, and D. Dominguez, Angular dependence of c-axis plasma frequency and critical current in Josephson-coupled superconductors at high fields, Phys. Rev. B, 1996, V. 53, P. 6634.

14 A. Pruymboom, P. H. Kes, E. van der Drift and S. Radelaar, Flux-Line Shear through Narrow Constraints in Superconducting Films, Phys. Rev. Letters, 1988, V. 60, P. 1430-1433.

15 S. Anders, A. W. Smith, R. Besseling, P. H. Kes, and H. M. Jaeger, Static and dynamic shear response in ultrathin layers of vortex matter, Phys. Rev. B, 2000, V.62, P. 15195-15199.

16 P. H. Kes, J. Aarts, V. M. Vinokur, and C. J. van der Beck, Dissipation in Highly Anisotropic Superconductors, Phys. Rev. Letters, 1990, V.64, P. 1063-1066.

17 R. Besseling, R. Niggebrugge, and P. H. Kes, Transport Properties of Vortices in Easy Flow Channels: A Frenkel-Kontorova Study, Phys. Rev. Letters, 1999, V.82, P. 3144-3147.

18 J. M. Harris, N. P. Ong, and Y. F. Yan, Hall Effect of Vortices Parallel to CuO2 Layers and the Origin of the Negative Hall Anomaly in YBa$_2$Cu$_3$O$_7$-delta, Phys. Rev. Letters, 1993, V.71, P. 1455-1458.

19 E. H. da Silva Neto, R. Comin, F. He, R. Sutarto, Y. Jiang, R. L. Greene, G. Sawatzky, A. Damascelli, Charge ordering in electron-doped superconductor Nd$_{2-x}$Ce$_x$CuO$_4$, Science, 2015, V. 347, P. 282-285.

20 P. Fournier, 'T'and infinite-layer electron-doped cuprates, Physica C: Supercond. Appl., 2015, V. 514, P. 314-338.

21 A. S. Klepikova, D. S. Petukhov, O. E. Petukhova, T. B. Charikova, N. G. Shelushinina and A. A. Ivanov, Incoherent interlayer transport in single-crystal films of Nd$_{2-x}$Ce$_x$CuO$_4$/SrTiO$_3$, Journal of Physics: Conference Series, 2018, V. 993, P. 12002-12007.

22 A. S. Klepikova, T. B. Charikova, N. G. Shelushinina, D. S. Petukhov and A. A. Ivanov, Anisotropy of the Hall Effect in a Quasi-Two-Dimensional Electron-Doped Nd$_{2-x}$Ce$_x$CuO$_{4+\delta}$ Superconductor, Physics of the Solid State, 2018, V. 60, P. 2162-2165.

23 A. Guarino, L. Parlato, C. Bonavolontá, M. Valentino, C. de Lisio, A. Leo, G. Grimaldi, S. Pace, G. Pepe, A. Vecchione and A. Nigro, Transport and optical properties of epitaxial Nd$_{1.83}$Ce$_{0.17}$CuO$_{4-\delta}$ thin films, J. Phys.: Conf. Series, 2014, V. 507, P. 012018.

24 A. Guarino, N.Martucciello, P. Romano, A. Leo, D. D'Agostino, M. Caputo, F.Avitabile, A.Ubaldini, G. Grimaldi, A.Vecchione, F.Bobba, C.Attanasio, and A. Nigro, Nd$_{2-x}$Ce$_x$CuO$_{4\pm\delta}$ Ultrathin Films Crystalline Properties, IEEE Transactions on Applied Superconductivity, 2018, V. 28, N. 7, P. 7501404.

25 A. Guarino, A. Leo, A.Avellaa, F.Avitabile, N.Martucciello, G. Grimaldi, A. Romano, S. Pacea,, P. Romano, A. Nigro, Electrical transport properties of sputtered Nd$_{2-x}$Ce$_x$CuO$_{4\pm\delta}$ thin films, Physica B, 2018,V. 536, P. 742-746.

26 M. R. Popov, A. S. Klepikova, N. G. Shelushinina, A. A. Ivanov, T. B. Charikova, Interlayer Hall Effect in n-type doped high temperature superconductor Nd2-xCexCuO4+δ, Physica C, 2019, V. 566, P. 1353515.

27 J. Bardeen, M. J. Stephen, Theory of the Motion of Vortices in Superconductors, Phys. Rev. A, 1965, V. 140,P. 1197.

28 N. G. Shelushinina, G. I. Harus, T. B. Charikova, D. S. Petukhov, O. E. Petukhova, A. A. Ivanov, The mixed-state Hall conductivity of single-crystal films Nd$_{2-x}$Ce$_x$CuO$_{4-d}$ (x = 0.14), Low temperature physics, 2017, V. 43, P. 475.

29 D. I. Khomskii and A. Freimuth, Charged Vortices in High Temperature Superconductors, Phys. Rev. Lett., 1995, V.75, P. 1384.

30 M. V. Feigel'man, V. B. Geshkenbein, A. I. Larkin and V. M. Vinokur, Sign Change of the Flux Flow Hall Effect in HTSC, JETP Lett., 1995, V.62, P. 834.

31 A. van Otterlo, M. Feigel'man, V. Geshkenbein and G. Blatter, Vortex Dynamics and the Hall Anomaly: A Microscopic Analysis, Phys.Rev. Lett., 1995, V.75, P. 3736.

32 X. X. Zhuo, Li Z. Feng, Yi X. Lei, F. J. Jia, Xu C. Qiang, Z. Nan, M. Yan, Z. Yu Feng, P. Y. Qiang, Q. L. Yao, Z. Wei, Z. H. Jun, S. Z. Xiang, Thermally activated flux flow, vortex-glass phase transition and the mixed-state Hall effect in 112-type iron pnictide superconductors, Science China Physics, Mechanics & Astronomy, 2019, V. 61, P. 127406.

33 N. B. Kopnin, B. I. Ivlev, V. A. Kalatsky, The flux-flow hall effect in type II superconductors. An explanation of the sign reversal, Journal of Low Temperature Physics, 1993, V. 90, P. 1–13.

34 N. B. Kopnin, Hall effect in moderately clean superconductors and the transverse force on a moving vortex, Phys. Rev. B, 1996, V. 54, P. 9475-9483.

35 A. G. Aronov, S. Hikami, and A. I. Larkin, Gauge invariance and transport properties in superconductors above $T_c$, Phys. Rev. B, 1995, V.51, P. 3880.

36 V.M.Genkin, A.S.Melnikov, Motion of Abrikosov vortices in anisotropic superconductors, JETP, 1989, V. 95, P. 2170-2174.

37 L. P.Gor'kov, N. B. Kopnin, Vortex motion and resistivity of type-ll superconductors in a magnetic field, Sov. Phys. Usp., 1975, V. 18, P. 496–513.

38 K. C. Woo, K. E. Gray, R. T. Kampwirth, J. H. Kang, S. J. Stein, R. East, and D. M. McKay, Lorentz-Force Independence of Resistance Tails for High-Temperature Superconductors in Magnetic Fields near Tc, Phys. Rev. Lett., 1989, V. 63, P. 1877-1879.

39 I. G. Gorlova, Yu. I. Latishev, The equivalence of the influence of a weak magnetic field and current on the resistance of single crystals $Bi_2Sr_2CaCu_2O_x$ is lower than the Berezinskii-Kosterlitz-Thouless transition temperature, JETP Letters, 1990, V. 51, P. 224-227.

40 W. Kohno, H. Ueki, and T. Kita, Hall Effect in the Abrikosov Lattice of Type-II Superconductors, Journal of the Physical Society of Japan, 2016, V. 85, P. 083705.

41 G. Venditti, J. Biscaras, S. Hurand, N. Bergeal, J. Lesueur, A. Dogra, R. C. Budhani, M. Mondal, J.Jesudasan, P. Raychaudhuri, S. Caprara, and L. Benfatto, Nonlinear I-V characteristics of two-dimensional superconductors: Berezinskii-Kosterlitz-Thouless physics versus inhomogeneity, Phys. Rev. B., 2019, V. 100, P. 064506.